\documentclass[preprint]{aastex}

\usepackage{color,amsmath}
\defcitealias{sok94}{S94}
\defcitealias{kim12}{KT12}
\defcitealias{mau06}{MH06}
\newcommand\oops[1]{{#1}}%
\newcommand{\vct}[1]{\vec{\mathbf{#1}}}
\newcommand{\cs}{c_\mathrm{s}}
\newcommand{\Varm}{V_{\rm pattern}}
\newcommand{\rmd}{\mathrm{d}}
\newcommand{\kmps}{km\,s${}^{-1}$}
\newcommand{\Msun}{$M_\sun$}
\newcommand{\Mspy}{\Msun\,yr${}^{-1}$}
\newcommand{\mach}{\mathcal{M}}

\shorttitle{BINARY EFFECTS ON AGB ENVELOPES}
\shortauthors{KIM \& Taam}

\begin{document}
\title{Wide Binary Effects on Asymmetries in Asymptotic Giant Branch 
Circumstellar Envelopes}
\author{Hyosun Kim\altaffilmark{1} and Ronald E. Taam\altaffilmark{1,2}}
\altaffiltext{1}{Academia Sinica Institute of Astronomy and Astrophysics, 
  P.O. Box 23-141, Taipei 10617, Taiwan; hkim@asiaa.sinica.edu.tw}
\altaffiltext{2}{Department of Physics and Astronomy, Northwestern University,
  2131 Tech Drive, Evanston, IL 60208; r-taam@northwestern.edu}

\begin{abstract}
Observations of increasingly higher spatial resolution reveal the 
existence of asymmetries in the circumstellar envelopes of a small 
fraction of asymptotic giant branch (AGB) stars. Although there is 
no general consensus for their origin, a binary companion star may be 
responsible. Within this framework, we investigate the gravitational 
effects associated with a sufficiently wide binary system, where 
Roche lobe overflow is unimportant, on the outflowing envelopes 
of AGB stars using three dimensional hydrodynamic simulations. The 
effects due to individual binary components are separately studied, 
enabling investigation of the stellar and circumstellar characteristics 
in detail. The reflex motion of the AGB star alters the wind velocity 
distribution, thereby, determining the overall shape of the outflowing 
envelope. On the other hand, the interaction of the companion with the 
envelope produces a gravitational wake, which exhibits a vertically 
thinner shape. The two patterns overlap and form clumpy structures. To 
illustrate the diversity of shapes, we present the numerical results as 
a function of inclination angle. Not only is spiral structure produced by 
the binary interaction, but arc patterns are also found that represent 
the former structure when viewed at different inclinations. The arcs 
reveal a systematic shift of their centers of curvature for cases when 
the orbital speed of the AGB star is comparable to its wind speed. 
They take on the shape of a peanut for inclinations nearly edge-on. 
In the limit of slow orbital motion of the AGB star relative to the 
wind speed, the arc pattern becomes nearly spherically symmetric. 
We find that the aspect ratio of the overall oblate shape of the 
pattern is an important diagnostic probe of the binary as it can 
be used to constrain the orbital velocity of the AGB star, and 
moreover the binary mass ratio.
\end{abstract}


\section{INTRODUCTION}\label{sec:intro}

Stars coevolve with their companion in binary (or multiple) systems. 
On the main-sequence, about 50\% of solar-type stars possess companions,
and their orbital separations show a log-normal distribution from 0.01 
AU to $10^4$ AU with the peak at about 30 AU, or a binary period of 300 
years \citep{rag10}. For evolved stars particularly in the asymptotic 
giant branch (AGB) phase, however, the significant extinction due to
the dense circumstellar envelopes makes the direct detection of the 
companions difficult. After ejection of the envelope, a fraction of 
stars in the post-AGB and planetary nebula (PN) phases reveal evidence
of binarities, although the samples are biased to the optically bright 
objects without strong stellar pulsations \citep[see][and references
therein]{van03,dem09}. The number of known binary systems in the phase
from AGB to PN is yet far from being statistically meaningful to be
compared with the binary fraction and orbital period of stars on the 
main-sequence.

Binaries are believed to play a crucial role in the stellar evolution
beyond the main-sequence \citep{hug07,dem09}. Of particular interest 
is the companion's role in facilitating the shape transition from 
AGB to PN phases. The circumstellar envelopes of AGB stars were often 
observed (and assumed) to be spherical \citep[e.g.,][]{ner98}, while 
the observed PNs show a stunning variety of shapes including jets 
and tori \citep[e.g.,][]{bal02,dem09}. In order to explain the shape 
transformation from AGB to PN, several scenarios including magnetic 
winds \citep[e.g.,][]{nor06,pas10}, mergers of stellar and substellar 
companions \citep{sok96,nor06}, and wind accretion around binary 
stars \citep[e.g.,][]{mor87,the93,dev09,moh11} have been suggested. 
Interestingly, most of the plausible scenarios favor the existence of 
companions \citep{hug07} and observationally the connection between
most of the PNs and bipolar morphologies is found \citep{mis09}.

The origin and onset of the aspherical geometry remains one of 
the fundamental questions in stellar evolution. A small number 
of AGB and post-AGB circumstellar envelopes have revealed their 
aspherical morphologies in spectropolarimetric observations 
\citep{joh91,tra94} and in the nascent bipolar reflection nebulae 
in near-infrared polarization imagings \citep[e.g.,][]{sch02}.
Some asymmetric fluctuation of the wind are traced by higher 
angular resolution observations of molecular line transition 
emission \citep[e.g.,][with order of 1\arcsec\ resolution]{cas10}.
With the subarcsecond resolutions of the Extended Very Large 
Array and the Atacama Large Millimeter/submillimeter Array, it 
is anticipated that a larger sample of AGB stars with asymmetric 
circumstellar envelopes will be found.

The central region of a binary system could be complex, depending
on the binary separation and the size of wind acceleration zone.
Close binary systems in which the AGB star fills its Roche lobe may 
encounter the common envelope phase \citep[e.g.,][]{ibe93,taa00}. 
In the cases of low mass companions, spin-orbit tidal interaction 
will lead to its migration and spiral in \citep[e.g.,][]{dar79,
nor10}, leading to the common envelope phase. In wider binary 
systems that avoid the common envelope interaction, the wind 
may be affected by the gravitational potential of the AGB star 
before reaching the Roche lobe \citep{moh11} if the radiative 
acceleration is insufficient to ensure escape, possibly occurring 
in oxygen-rich outflows \citep{win00}. Studying the innermost 
part of the wind is important for understanding the direct binary 
interaction, however there are many difficulties. First, most of 
the AGB circumstellar envelopes are extremely dense because of 
the substantial rates of mass loss ($10^{-7}$--$10^{-4}$\,\Mspy). 
As a result, the direct detection of the binary companion is 
challenging. Second, to investigate the gas accretion process, 
high spatial resolution for line transition observations is 
required. The highest resolution is about $0.\!\!\arcsec1$, 
corresponding to 100\,AU scale at a distance of 1\,kpc.
At 100\,AU, the escape velocity from a 1\,\Msun\ star is about 
4\,\kmps, and thus a detectable binary system of this separation
is unlikely to experience mass transfer via Roche lobe overflow
unless the wind is slower than or comparable to 4\,\kmps. Since
oxygen-rich winds are typically slow \citep[e.g., Mira]{win00,
fon06}, it may be possible to detect the wind Roche lobe overflow 
\citep{moh11} with current high resolution observation. It will be 
necessary to extend the investigation of such objects beyond 1\,kpc 
in order to obtain a statistically meaningful sample of AGB binary 
systems to overcome the short AGB timescale ($\sim10^6$\,years).

On a larger scale, several theoretical and numerical works have 
shown that the orbital motion of binary stars deforms the geometry 
of the mass loss, producing a spiral pattern in the orbital plane 
\citep[e.g.,][]{sok94,mas99,he07,edg08,rag11}. Particularly important was 
the observational discovery of a well-defined Archimedes spiral pattern 
in the circumstellar envelope of AFGL 3068 in dust scattered light, 
strongly suggesting the existence of a companion to this evolved star 
\citep{mau06}. The binary scenario for this spiral pattern is strongly 
supported by a detection of two point sources in the near-infrared 
\citep{mor06}. Such a binary scenario has also been discussed as a 
possible explanation for the asymmetries emerging from the molecular 
line emission of CIT 6 \citep{din09} and the dust thermal emission 
in the far-infrared of Mira \citep{may11} and IRC+10216 \citep{dec11}.

The circumstellar spiral pattern imprinted over a large scale is of 
significant importance because it preserves the information of the 
binary hidden in the deep central part of the dense circumstellar 
envelope. Wide binary systems are of particular interest in testing
the binary-spiral scenario as their long orbital periods induce the 
spiral arm spacings sufficient to be resolved by current observational 
techniques. For close binary stars encountering the common-envelope 
interactions, the overall mass loss geometry on a larger scale is
likely affected by their complex evolution in the very central region. 
Nevertheless, the study of the spiral patterns in wide binary 
systems will facilitate the understanding of their binary systems.

To interpret the large spacing observed in the circumstellar patterns of 
AGB and post-AGB stars (e.g., 2300 AU in AFGL 3068 for an assumed distance 
of 1 kpc), we focus on a wide binary system, where the system is detached 
and Roche lobe overflow is unimportant. In a wide comparable-mass binary 
system, the companion star influences the mass outflow in two ways. The 
companion focuses a fraction of the wind material via its gravitational 
potential, building a gravitational density wake \citep[\citetalias{kim12} 
hereafter]{kim11,kim12} similar to the Bondi-Hoyle-Lyttleton process 
\citep{hoy39,bon44,bon52}. In addition, the companion indirectly affects 
the geometry of the wind by forcing the AGB star to move about the center 
of mass of the system, thereby breaking the spherical symmetry of the 
outflowing wind \citep[e.g.,][\citetalias{sok94} hereafter]{sok94}.

In order to determine the anisotropy of the wind due to the reflex 
motion of the mass losing star, \citetalias{sok94} performed a 
coordinate transformation from the rest frame of the star to the 
center of mass frame of the system. \citetalias{sok94} found the 
conversion factor between the solid angles in these two coordinate 
systems, which modifies the mass loss geometry. \citet{he07} extended 
this work and examined the dependence of the column density patterns 
on model parameters. A simple piston model, which provides a reasonable 
approximation for the structure at large distances, was employed 
to increase the computational efficiency. Various shapes were shown 
which depended on the viewing angle and the eccentricity of the binary 
orbit. In addition, a qualitative description for the strength of the 
pattern density was presented in terms of the ratio between the wind 
and orbital velocities. A quantitative investigation is needed to 
provide an interpretative framework for the observed AGB patterns, 
taking into account the effect of the masses of both stars and the 
gas pressure of the wind material (ignored in the models 
of \citetalias{sok94} and \citealp{he07}).

\citet{mas99} investigated smoothed particle hydrodynamic simulations 
of such systems taking into account the gravitational influences of 
the binary stars, a fictitious force to accelerate the gas in the dust 
formation zone, and heating and cooling mechanisms. This seminal work 
was especially prescient in showing the role of evolved stars in binary 
systems in producing a spiral shock pattern. 
Close binary cases in which the wind is not efficiently accelerated to 
the escape velocity were also investigated, which revealed significant 
focusing of the wind material onto the orbital plane, mimicking bipolar 
outflows, in addition to the spiral pattern in the orbital plane. 
However, their inclusion of complex physical processes in the circumstellar 
envelope obscured the main physical reason of the various pattern shapes. 

In a subsequent study, \citet{edg08} used a grid based code and 
reduced the complexities of the input physics, primarily focusing 
on the effects associated with the binary motions. By adjusting 
the gravitational softening radius of the companion, it was found 
that the overall density structure arises from the reflex motion 
of the mass losing star, but that the spiral shock temperature is 
governed by the companion. Considering that the wind speed adopted 
in their simulations is extremely fast (38 and 65\,\kmps) compared 
to the reflex motion of the mass losing star (of order 3\,\kmps), 
the effect of reflex motion is considerably suppressed. In contrast, 
the density contrast of the companion's gravitational wake does not 
strongly depend on the wind velocity in the fast wind regime 
\citepalias[see Equation (17) of][]{kim12}.

Here, we separately examine the effects of the orbital motions 
of the individual stars in a binary system on the structure of 
the outflowing envelope, enabling us to investigate the binary 
stellar and circumstellar properties in detail. We first focus 
on the hydrodynamic effects associated with the reflex motion 
of the AGB star in detail, by suppressing the direct influence 
of companion's gravitational potential. With an understanding 
of the mass flow variation due to the reflex motion, we then 
include the gravitational wake of the companion. In a previous 
paper \citepalias{kim12}, we investigated the gravitational wake 
of a low mass companion such as a planet or a brown dwarf, which 
does not induce the AGB star to revolve about the center of mass 
of the system. We will show that the results in \citetalias{kim12} 
can be applied to a system characterized by components of comparable 
mass.

This paper is organized as follows. In Section \ref{sec:analy}, we 
first revisit the analytical study of \citetalias{sok94} in order 
to obtain the magnitude of the density and velocity fluctuations 
in the mass loss geometry without considering hydrodynamic effects. 
In Section \ref{sec:setup}, we describe the numerical setup for the 
three dimensional hydrodynamic simulations, defining the symbols and 
terms used in this paper. In Section \ref{sec:resul}, we present our 
findings from the results of the hydrodynamic simulations studying the 
wind velocity (Section \ref{sec:vwind}), the density profile (Section 
\ref{sec:alpha}), the morphology (Section \ref{sec:morph}), and the 
effects of companion's gravitational wake (Section \ref{sec:2arms}).
In Section \ref{sec:discu}, we apply our model to constrain the binary 
orbital properties of a carbon-rich binary system, AFGL 3068. 
We summarize and conclude in Section \ref{sec:concl}.

\section{GEOMETRICAL EFFECTS OF REFLEX MOTION OF MASS LOSING STAR: 
ANALYTIC ANALYSIS}\label{sec:analy}

\begin{figure}
  \epsscale{0.4}
  \plotone{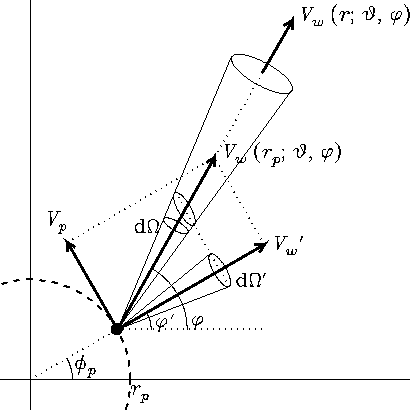}
  \caption{\label{fig:sktch}
    Schematic diagram illustrating the modification of wind geometry due to 
    the orbital motion of the mass losing star in a binary system. The mass 
    losing star is located at $(r,\,\theta,\,\phi)=(r_p,\,\pi/2,\,\phi_p)$,
    marked by a filled circle, and orbiting on the dashed circle. Its wind 
    velocity $\vct{V}_w$ in the center of mass frame is determined by the 
    vector sum of the intrinsic wind velocity $\vct{V}_w^\prime$, which is 
    the velocity in the frame of the mass losing star, with the orbital 
    velocity of the star $\vct{V}_p$. The solid angle of a wind parcel 
    is consequentially modified to $\rmd\Omega=S\rmd\Omega^\prime$. See
    text \citepalias[and][]{sok94} for details of the projection factor 
    $S$. The directions of velocity vectors are defined by the spherical 
    coordinate angles $\vartheta$ and $\varphi$ about the position of the 
    mass losing star.}
\end{figure}

We start by describing the geometrical effects of the orbital motion 
of a mass losing star on the density and velocity of its circumstellar 
envelope. Figure\,\ref{fig:sktch} is a schematic diagram displaying 
the geometrical change of a velocity element in the stellar wind. The 
stellar mass loss is assumed to be intrinsically isotropic with a wind 
velocity of $\vct{V}_w^\prime$, radially expanding in the rest frame of 
the mass losing star. Under the influence of the gravitational potential 
of the secondary star, the orbital reflex motion of the mass losing star 
alters the direction and magnitude of the wind velocity in the center of 
mass frame to $\vct{V}_w=\vct{V}_w^\prime+\vct{V}_p$, where $\vct{V}_p$ 
is the orbital velocity. In order to indicate the velocity directions and 
to avoid confusion with the observation position $(r,\,\theta,\,\phi)$, 
we define the spherical angles relevant to the location of the mass losing 
star as $\vartheta$ and $\varphi$. Thus, the directions for the intrinsic 
wind velocity $\vct{V}_w^\prime$ and the wind velocity in the center of mass 
frame $\vct{V}_w$ are denoted by $(\vartheta^\prime,\,\varphi^\prime)$ and 
$(\vartheta,\,\varphi)$, respectively. As shown in Figure\,\ref{fig:sktch},
as an example, the intrinsic velocity of a wind parcel $\vct{V}_w^\prime$ 
perpendicular to the orbital velocity $\vct{V}_p$ in the orbital plane 
($\vartheta^\prime=\pi/2$; $\varphi^\prime=\phi_p$) corresponds to the 
wind velocity observed in the center of mass frame having the magnitude 
of $V_w=(V_w^{\prime\,2}+V_p^2)^{1/2}$. 

As a result, the solid angle of a wind parcel is consequently 
modified. The wind parcel ejected from the star in the solid 
angle $\rmd\Omega^\prime$ is shifted laterally due to the inertia 
of the flow with the orbital velocity of the mass losing star. 
The projected area normal to the direction of $\vct{V}_w$ is 
smaller by a factor of $(V_w^\prime/V_w)$. From the definition 
of a solid angle, the area at a unit distance from the star is
multiplied by $(V_w^\prime/V_w)^2$. Hence, the solid angles between 
the intrinsic wind ($\rmd\Omega^\prime$) and the wind observed in 
the center of mass frame ($\rmd\Omega$) satisfy the relationship 
of $\rmd\Omega=(V_w^\prime/V_w)^3\,\rmd\Omega^\prime$ for the wind 
parcel intrinsically perpendicular to the orbital motion. We note 
that the orbital reflex motion of the mass losing star changes not 
only the magnitude of the wind velocity but also the mass loss 
geometry (i.e., mass loss rate $\dot{M}_p$ per solid angle). This 
leads to a density in the direction of $\vct{V}_w$ for the wind 
parcel in Figure\,\ref{fig:sktch} given as
\begin{equation}
  \rho_{\rm parcel}=\frac{1}{r^2V_w}\frac{\rmd\dot{M}_p}{\rmd\Omega}
  =\left(\frac{V_w}{V_w^\prime}\right)^2\frac{\dot{M}_p}{4\pi r^2V_w^\prime},
\end{equation}
making the relatively faster wind parcel relatively denser\footnote{%
Strictly speaking, the distance $r$ should be replaced by the distance 
from the star, $(r^2+r_p^2-2r_p r\sin\theta\cos(\phi-\phi_p))^{1/2}$, 
where $r_p$ and $\phi_p$ refer the orbital radius of the mass losing 
star and the azimuthal angle in the orbit, respectively.}. Compared 
to the density distribution around the corresponding single star, 
$\rho^\prime=\dot{M}_p/(4\pi r^2V_w^\prime)$, the contribution of 
this wind parcel emerging from the star with the reflex motion to 
the density enhancement ratio, $\rho_{\rm parcel}/\rho^\prime-1$, 
is $(V_p/V_w^{\prime})^2$. Similarly, simple geometrical analysis 
leads to the result that the wind parcels in the forward and 
backward directions ($\varphi^\prime=\phi_p\pm\,\pi/2$) have maximum 
and minimum values for wind matter in the center of mass frame for 
velocities ($V_w=V_w^\prime\pm V_p$) and density enhancement ratios 
($\pm\,V_p/V_w^\prime$), respectively.

To use \citetalias{sok94}'s analytic model as a basis for the 
analysis of our hydrodynamic simulations, we relax his linear 
assumption ($V_p\ll V_w^\prime$) and fully derive the modulation 
of the velocity and density of material released from the AGB 
star in its orbital motion. 
However, this simple analysis is used to track a parcel in each 
direction, but does not take account of the interaction between 
the parcels emerging from the star at different positions in the 
orbital motion. Our numerical simulation indeed exhibits the 
departures from the prediction based on this simple analysis.

\section{NUMERICAL METHOD}\label{sec:setup}

The FLASH3 code is used to solve the governing equations of hydrodynamics
\begin{equation}\label{equ:conti}
  \frac{\partial\rho}{\partial t}+\vct{\nabla}\cdot(\rho\vct{V})=0,
\end{equation}
and
\begin{equation}\label{equ:momen}
  \frac{\partial\vct{V}}{\partial t}+\vct{V}\cdot\vct{\nabla}\vct{V}
  = -\frac{\cs^2}{\rho}\vct{\nabla}\rho-\vct{\nabla}\Phi_p^{\rm eff}
  -\vct{\nabla}\Phi_{\rm comp}.
\end{equation}
where we assume an isothermal flow with sound speed $\cs$, which 
is reasonable for an AGB circumstellar envelopes beyond the wind 
acceleration zone. Here, $\Phi_{\rm comp}$ is the gravitational 
potential of the companion in a circular motion. The companion is 
assumed to be a main-sequence star without mass loss. The effective 
potential $\Phi_p^{\rm eff}$ represents the gravitational potential 
of the mass losing star $-GM_p/r$, but it also includes a term $f/r$ 
to approximate the effect of stellar radiation pressure on dust as a 
wind acceleration mechanism. We have carried out several simulations 
with different values of $f$ to clarify that the wind acceleration does 
not affect our results as far as the wind can escape. For the models 
presented in this paper (Table\,\ref{tab:param}), $f=1$ is employed so 
that the gas motion is only governed by the boundary condition at 
the surface of the mass losing star, the gravitational field of the 
companion, and the pressure gradient in the flow. Alternatively, this 
setup can be interpreted as that the wind is fully accelerated at the 
numerical surface of the mass losing star (2 AU from the star). In 
Table\,\ref{tab:param}, all parameters with a subscript `$p$' denote 
the quantities related to the mass losing star, e.g., the orbital 
radius of the mass losing star is $r_p$. The quantities for the 
companion star are denoted by a subscript `comp'. The subscript 
`$w$' refers to the stellar wind mass loss from the primary star.

In order to compare the resulting structure with the corresponding 
single star case, we define the background as follows. We have carried 
out the same simulation with the individual model by placing the mass 
losing star at distance $r_p$ from the center, but without introducing 
the orbital motion. The background density $\rho_0(r)$ and velocity 
$V_{w,0}(r)$ are defined as the averaged values over the azimuthal 
direction at distance $r$ for the steady distribution about the single 
star located at $r_p$. By calculating the average values, we match 
the position of the star between the cases that the star is single 
and in a binary system. The notations with a subscript `0' defined 
as above are technically different from \citetalias{sok94}'s notations 
with a superscript `$\prime$' as measured in the rest frame of the mass 
losing star. However, the differences are negligible when the pattern
of interest is located at a large distance from the star.

We first examine the anisotropic properties of wind material (or 
circumstellar envelope) purely due to the reflex motion of the 
mass losing star, by excluding the direct influence of companion's 
gravitational potential (i.e., $\vct{\nabla}\Phi_{\rm comp}=0$) 
in most of models (M1--M6 in Table\,\ref{tab:param}). For these 
models, only the indirect contribution of the companion causing 
the motion of the mass losing star is taken into account, which is 
analogous to \citetalias{sok94} except for the inclusion of gas 
pressure and parcel interaction in the present study. We include 
the companion's gravitational potential in model M7 to compare with 
results of model M6. The interpretation of the results from model M7 
are based on the results found from models M1--M6 as well as results 
based on patterns solely due to the companion (gravitational density 
wake; see \citetalias{kim12}).

\section{RESULTS}\label{sec:resul}

\begin{figure}
  \plotone{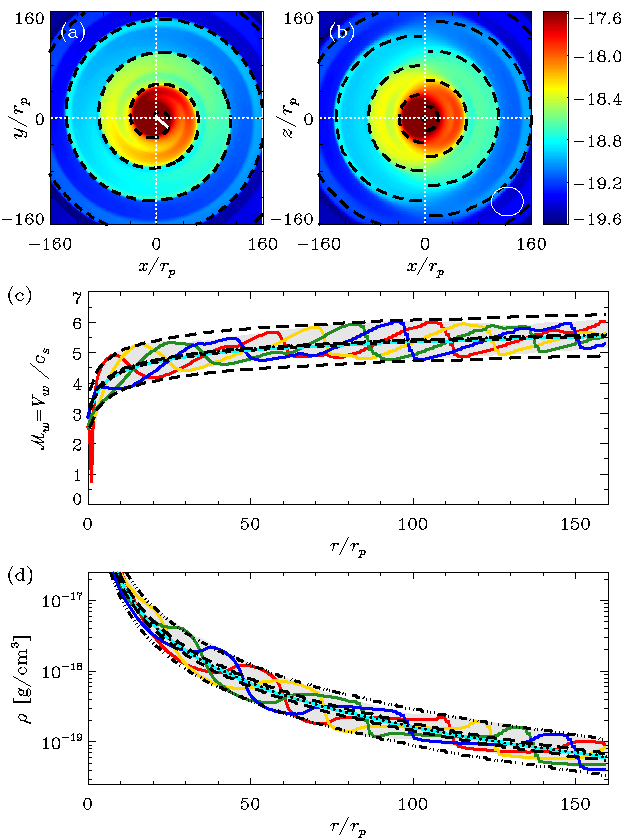}
  \caption{\label{fig:subso}
    Model M1 for a subsonic orbital motion ($\mach_p=0.8$) of the mass 
    losing star. The density distributions (a) in the orbital plane 
    and (b) in a meridional plane are coded in a logarithmic color scale 
    as labeled in the color bar. The dashed lines outlining the spiral
    pattern in (a) and the circular arc pattern in (b) are drawn based
    on the empirical formulae in Section \ref{sec:speed}. In (a), the 
    white straight line from the center shows the standoff distance of
    the spiral pattern, described in Section \ref{sec:stand}. At the
    bottom right corner of (b), the ellipse displays the overall shape
    in meridional planes (see Section \ref{sec:vflat}). (c) The total
    velocity profiles of wind flow in the unit of sound speed $\cs$ as 
    a function of distance from the orbital center in $^+x$ (red), $^-y$ 
    (yellow), $^-x$ (green), and $^+y$ (blue) directions show variation 
    between $\mach_{w,0}(1\pm\beta_0)$ (top and bottom long dashed), 
    where $\mach_{w,0}$ and $\mach_p$ refer the Mach numbers of the 
    intrinsic wind motion (dotted) and the orbital motion, respectively. 
    Middle long dashed line displays the Mach number averaged over 
    azimuthal direction. The wind flowing along the orbital axis (cyan) 
    has the predicted speed $\mach_w(z)=\mach_{w,0}(1-\beta_0^2)^{1/2}$ 
    (short dashed) where $\beta_0=\mach_p/(\mach_{w,0}+1)$. (d) The 
    corresponding density profiles (red, yellow, green, blue, and cyan 
    colors) are located between the dot-dot-dot-dashed lines indicating 
    $\rho=\rho_0(1+\mach_p)^{\pm1}$, where $\rho_0$ (dotted) fits with 
    the average density profile. For comparison, long dashed lines show 
    the profiles for $\rho=\rho_0(1+\beta_0)^{\pm1}$. All the lengths are 
    scaled by the orbital radius $r_p$ of the mass losing star.
  }
\end{figure}

To provide an overview of the structures resulting from the hydrodynamic 
simulations, a comparison of two models corresponding to different orbital 
speeds are presented. The cases for subsonic ($\mach_p=0.8$; model M1) and 
supersonic ($\mach_p=3.0$; model M2) orbital motion of the mass losing star 
are illustrated in Figures\,\ref{fig:subso} and \ref{fig:super}, respectively. 
Individual panels show (a) a single-armed spiral in the orbital plane, 
(b) circular arcs in a meridional plane, (c) fluctuations in the velocity 
profile along the distance from the system center, and (d) deviations of 
the density profile from the background $\sim r^{-2}$ density distribution.

A slow orbital motion of the mass losing star is expected to only 
slightly modify the gas flow pattern as a given parcel of the wind 
would not interact with other parcels. This suggests that the resultant 
wind properties would be analogous to \citetalias{sok94}'s diagnostics, 
except for the effect of gas pressure. Figure \ref{fig:subso}(c) indeed 
shows that the wind velocity varies between $V_{w,0}(1\pm\beta_0)$ 
(long dashed lines) confirming the previous work by \citetalias{sok94}%
\footnote{The definition of $\beta_0=V_p/(V_{w,0}+\cs)$ in this paper 
is slightly different from \citetalias{sok94}'s definition for 
$\beta^\prime=V_p/V_w^\prime$ without $\cs$. See Section \ref{sec:vwind}.}. 
The density profile in Figure \ref{fig:subso}(d) smoothly varies with 
a constant enhancement ratio (limited by the dot-dot-dot dashed lines), 
retaining a description as a power law with distance $r$ from the orbital 
center (roughly $-2$) as $\rho_0$. 

In the case that the orbital motion is supersonic (model M2), 
however, the variations in the velocity and density profiles 
(Figure \ref{fig:super}(c)--(d)) significantly decrease with 
distance in the orbital plane. This is likely due to the fact 
that wind parcels with different velocities, emerging at different 
times, bunch up and create a shock. In particular, it is seen that 
the peak velocities at the shock fronts level off after the first 
shock encounter. 

\subsection{Wind Velocity}\label{sec:vwind}

Based on additional model computations for the subsonic motion of the 
mass losing star, it is found that the ratio of the orbital speed of 
the star to the intrinsic wind speed, as \citetalias{sok94} suggested, 
is a good indicator for the magnitude of the velocity fluctuation, 
$(V_w-V_{w,0})/V_{w,0}$. However, in supersonic cases, this velocity 
ratio is not the best indicator for the variation in velocity profile. 
Specifically, the velocity for the wind parcel moving along the $z$-axis 
shows approximately constant excess as a function of distance as compared 
to \citetalias{sok94}'s expectation\footnote{In \citetalias{sok94}, the
orbital motion of the mass losing star modifies the velocity of the wind 
parcel moving along the $z$-axis to be $V_w=V_w^\prime(1-\beta^{\prime2}
)^{1/2}$ where $\beta^\prime=V_p/V_w^\prime$.}. 

We find that a revised velocity 
ratio defined by $\beta_0=V_p/(V_{w,0}+\cs)$ matches the wind velocity 
for the parcel moving along the $z$-axis (cyan-colored line in Figure 
\ref{fig:super}(c)) when used with \citetalias{sok94}'s corresponding 
formula (short dashed line; $V_w(z)=V_{w,0}(1-\beta_0^2)^{1/2}$). 
Including the sound speed term in the definition of $\beta_0$ reduces 
the mean absolute deviation of the velocity difference from the formula 
for the wind parcel along the $z$-axis to 0.013\% and 2.6\% for M1 and 
M2 models, which are considerably smaller than the corresponding values 
without $\cs$ (0.034\% and 9.7\%, respectively). For the subsonic model 
M1, the deviations from \citetalias{sok94}'s formulas for the azimuthal 
maximum and minimum velocities in the orbital plane are also reduced 
to 2.4\% and 5.1\% (from 4.9\% and 7.2\%).

In model M2 (Figure \ref{fig:super}(c)), the wind velocity at the 
first shock position matches with \citetalias{sok94}'s formula for 
the maximum wind velocity. However, after encountering the first 
shock, the variation in wind flow speed is reduced, i.e., the maximum 
speed at a distance is smaller, and the minimum speed is larger, than 
the prediction based on the simple analysis of \citetalias{sok94}.

We point out that the wind velocities vary in a binary system depending 
on the direction as shown in, for example, Figure \ref{fig:super}(c). 
Here, the maximum velocity is $\sim6.7$\,\kmps\ in the orbital plane, but 
$<4.5$\,\kmps\ along the orbital axis, implying that the line-of-sight 
expansion velocity (usually determined by half spectral line width of 
molecular line emission) does not necessarily determine the pattern 
spacing appearing in the plane of the sky. Furthermore, this velocity 
(after modulated by the orbital motion of the binary system) is larger 
than that in the intrinsic wind model (dotted line; $<5$\,\kmps), which 
could be considered as mimicking a wind acceleration mechanism.

Finally, the variation of the maximum velocity (see above) with latitude 
also implies that the spectral line width can differ between an edge-on 
and (high angular resolution) face-on observation. A low resolution 
face-on observation will have a spectral line width between the values 
expected from the maximum velocities of the wind elements flowing through 
the orbital plane and along the orbital axis, representing the projected 
velocities of the flows over the entire envelope. However, a high angular 
resolution observation using an interferometer can resolve the central 
part, which in the case of a face-on view, corresponds to the flow along 
the orbital axis. The differing spectral line width in the central part 
may indicate the flow variation due to the (undetected) binary companion.

\subsection{Density Profile}\label{sec:alpha}

\begin{figure}
  \epsscale{0.8}
  \plotone{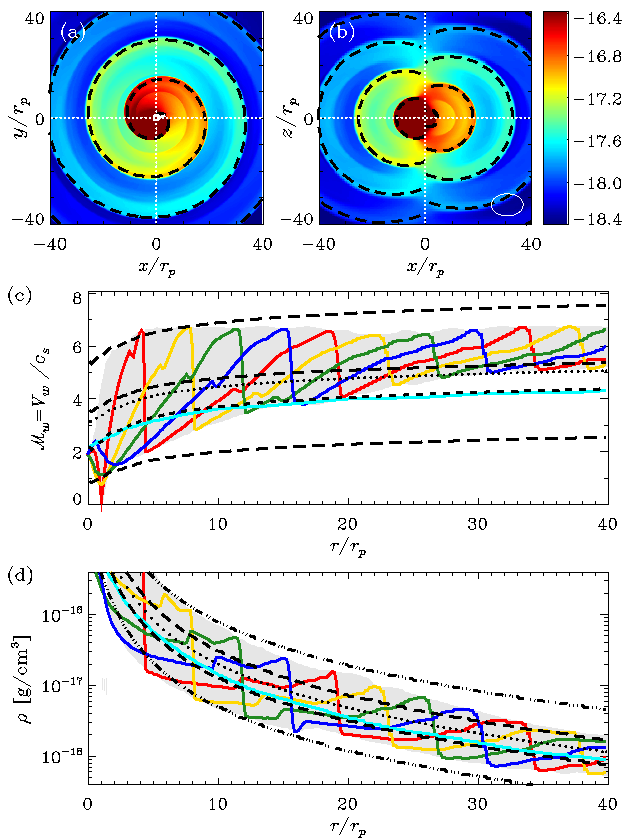}
  \caption{\label{fig:super}
    Model M2 for a supersonic orbital motion ($\mach_p=3.0$) of the mass 
    losing star. See Figure~\ref{fig:subso} for the details.
  }
\end{figure}

As the density contrast of the pattern can provide a diagnostic for 
the binary orbital parameters, we have illustrated in the last panel 
of each figure (\ref{fig:subso}(d) and \ref{fig:super}(d)), the 
density profile as a function of the distance from the center along 
$^+x$ (red), $^-y$ (yellow), $^-x$ (green), and $^+y$ (blue) in the 
orbital plane as well as along the orbital axis ($^\pm z$; cyan).

To quantify the overall density distribution as a function of distance, 
the average density profile is defined as the average value of the 
density over all directions in the orbital plane at each distance. From 
all the simulated models, we find that the average density profile is 
the same as the intrinsic density profile ($\rho_0$; dotted) in both 
slope and magnitude, except for a slight deviation in the central region.

The density profile along the orbital axis (cyan) also preserves 
the value of $\rho_0$ except in models M2 and M3, which have the 
largest value of $\beta_0$ (i.e., $V_p/(V_{w,0}+\cs)$). In these 
models with fast orbital motions in a slow wind, the circumstellar 
envelopes are rarefied along the axis with shallower profiles, 
relative to the isotropic density of the corresponding single mass 
losing star ($\rho_0$). We note that since $\rho_0$ corresponds to 
the average density profile in the orbital plane, an ideal edge-on 
observation may show a deviation in density profiles depending on
the direction angle from the orbital axis.

To quantify the maximum and minimum densities over all directions 
in the orbital plane, we make use of the numerical results displayed 
in Figure \ref{fig:subso}(d) for a subsonic orbital motion of the 
mass losing star (model M1). The density 
profiles along the five different directions (in color) are confined 
between $\rho_0(1+\mach_p)$ and $\rho_0(1+\mach_p)^{-1}$ (upper and 
lower dot-dot-dot dashed lines, respectively). The density contrast, 
$\rho_{\rm max}/\rho_{\rm min}$, is thus $(1+\mach_p)^2$ at all 
distances (i.e., 3.24 for model M1).

\begin{figure}
  \plotone{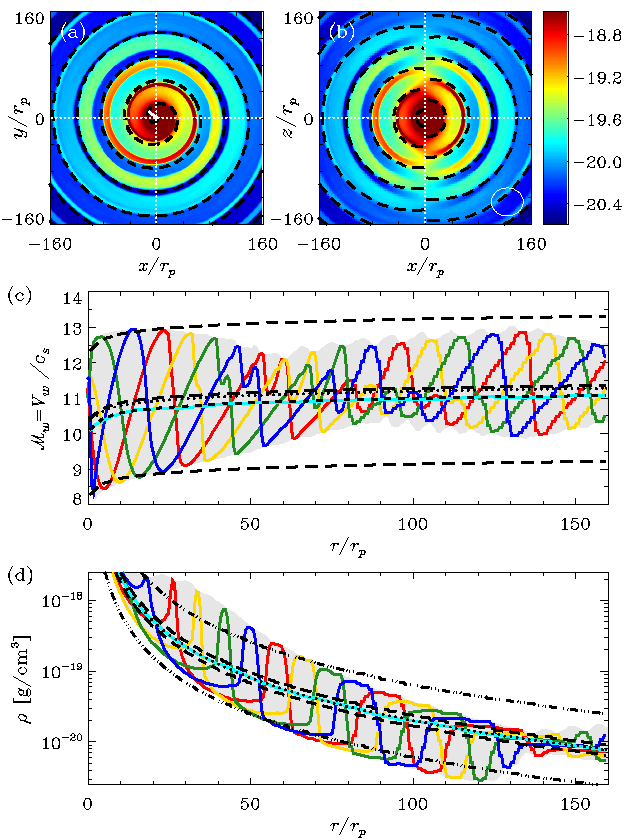}
  \caption{\label{fig:fastw}
    Same as Figures~\ref{fig:subso} and \ref{fig:super}, but for model 
    M6 with a faster wind ($\mach_w\sim10$). See Figure~\ref{fig:subso} 
    for the details.
  }
\end{figure}

On the other hand, the density profiles for the case of supersonic 
orbital motion (model M2) are displayed in Figure \ref{fig:super}(d) 
showing significant deviation from the dot-dot-dot dashed curves. 
The deviation becomes larger with distance, particularly beyond 
$r/r_p\sim16$, reducing the density contrast. This decrease of density 
contrast is also seen in Figure \ref{fig:fastw}(d) for model M6 with 
a faster wind. This deviation is due to the peak densities forming 
a steeper profile than the average density profile (dotted), and the 
minimum density profile suddenly changes at $r/r_p\sim130$. We note 
that the maximum and minimum density profiles eventually attain values 
given by $\rho_0(1+\beta_0)$ and $\rho_0(1+\beta_0)^{-1}$, respectively 
(long dashed lines).

A characteristic of the density profile is the existence of double 
peak features, which is best illustrated in Figure \ref{fig:super}(d). 
For example, along $^-x$-axis (green line), the first arm extends 
between $r/r_p\sim8$ and 12 bounded by sharp edges as hydrodynamic 
shocks. The second arm is found to exist between $r/r_p\sim14.5$ and 
26.5, again bounded by sharp edges. By comparing these two arms, one 
concludes that the arm width significantly increases in this model. 
The position of the outer edge is drawn by dashed lines in Figure 
\ref{fig:super}(a)--(b), and the remaining density discontinuity 
traces the inner edge of the arm. Because of the shape (pitch angle) 
difference, the inner and outer edges overlap at around $^+y$-axis 
($y/r_p\sim16$), producing a thin sub-arm structure. The red line in 
Figure \ref{fig:super}(d) has a tiny bump at $r/r_p\sim19$ on top of 
the outer edge, and the green line shows a knee at $r/r_p\sim22$ (also 
seen in the corresponding velocity profiles).

In the model with a faster wind (Figure\,\ref{fig:fastw}), the double 
peaks are compressed by the fast wind so that they are not resolved as 
the density variations in the arm are small. However, the double peak 
feature is clearly seen in the velocity profiles. In particular, the 
red line in the density profile has a box shape with edges, for instance, 
at $r/r_p\sim80$ and 100, corresponding to peaks in the velocity profile 
at $r/r_p\sim80$ and 95. We note that the outer edge has a velocity peak 
at the density maximum, but that the inner velocity peak is located at 
the density minimum, which differentiates them.

The inner peaks in Figure \ref{fig:fastw}(c) disappear beyond 
$r/r_p\sim120$, where the minimum density values in Figure 
\ref{fig:fastw}(d) suddenly jump to $\rho_0(1+\beta_0)^{-1}$. 
A similar jump of density minima occurs in 
Figure \ref{fig:super}(d) for the model M2 with a slow wind, 
but the density minima beyond $r/r_p\sim16$ are still smaller 
than $\rho_0(1+\beta_0)^{-1}$.

Eventually, the profiles of density maxima and minima in the models 
with supersonic orbital motion become steeper interior to the distance 
for the first overlap between inner and outer edges of the arm 
(Table\,\ref{tab:dfits}), than for the density profile characteristic 
of a single mass losing star ($\sim r^{-2}$). We note that the average 
of the density over all azimuthal directions at a distance is nearly 
same with $\rho_0$, despite the steeper profiles for maxima and minima. 
Beyond the overlap distance, the extrema may follow $\rho=\rho_0
(1+\beta_0)^{\pm1}$ with $+$ and $-$ signs for the maxima and minima, 
respectively, resulting in a negligibly small density contrast.

\subsection{Morphology}\label{sec:morph}

\subsubsection{Pattern}\label{sec:speed}

The shape of the pattern in the outflowing wind is determined by 
the pattern propagation speed $\Varm$ in the radial direction. 
The spiral shape in the orbital plane can be expressed by the 
differential equation satisfying
\begin{equation}\label{eqn:shape}
  \frac{d}{d(-\phi)}\left(\frac{r}{r_p}\right)=\frac{\Varm}{V_p}.
\end{equation}
Hence, measuring $\Varm$ is key in connecting the shape of the 
observed pattern to the orbital velocity $V_p$ of the mass losing 
star. As shown in Figure \ref{fig:super}(b), high density regions 
are bounded by circular shapes in a meridional plane. The dashed 
partial circles denote the outer boundaries of the enhanced density 
regions. By measuring the difference in position between adjacent 
circles, the pattern propagation speed averaged over one orbital 
period can be deduced.

Upon inspection of Figure \ref{fig:super}(b), there is a systematic 
shift of the centers of the circles. This can be understood by simply 
tracing the locations of the wind parcels emerging at the same time. 
For a duration $\Delta t$, the individual wind parcels travel 
different distances, i.e., $(V_w^\prime+V_p)\Delta t$ in the forward 
direction, $(V_w^\prime-V_p)\Delta t$ in the backward direction, and 
$(V_w^{\prime2}-V_p^2)^{1/2}\Delta t$ in the lateral direction. Thus, 
the emerging parcels describe a circle about the center shifted by 
$V_p\Delta t$ with a radius of $V_w^\prime\Delta t$ in a meridional 
plane. The resulting spiral pattern emerging in the orbital plane has
an interval between arms given as $(V_w^\prime+V_p)\times T_p$, where 
$T_p$ is the orbital period. We note that this is only approximate 
because (1) a coordinate transformation is required from the frame 
comoving with the mass losing star to the observer's rest frame, (2) 
piling wind parcels with different velocities occurs, (3) $V_{w,0}$ 
is not exactly same with $V_w^\prime$, and (4) gas pressure can affect 
the final configuration.

\begin{figure}
  \epsscale{1}
  \plotone{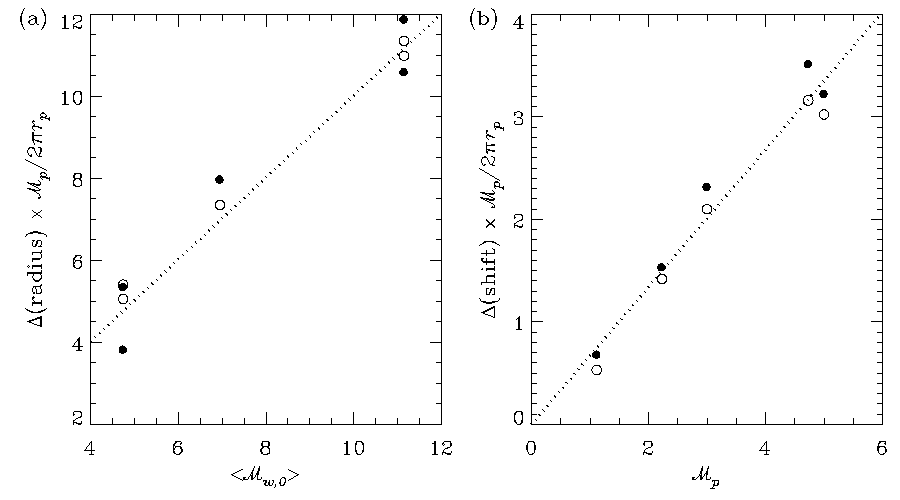}
  \caption{\label{fig:speed}
    The pattern propagation speed for M2--M6 models in Mach number in 
    the radial direction decomposed to two parts, $\mach_{\rm radius}$ 
    and $\mach_{\rm shift}$, relevant to the circular shapes found in 
    meridional planes. (a) $\mach_{\rm out,radius}$ (open symbols) and 
    $\mach_{\rm int,\,radius}+2$ (filled symbols) versus the average wind 
    Mach numbers, $<$$\mach_{w,0}$$>$, follow the dotted line of slope 
    1. The linear fit for the slope of the 10 points is $1.04\pm0.03$. 
    (b) The centers of the circles shift with the speed proportional 
    to the orbital speed. $\mach_{\rm out,shift}$ (open symbols) and 
    $\mach_{\rm int,\,shift}+0.5$ (filled symbols) as a function of 
    the orbital Mach number, $\mach_p$, are linearly fitted with the 
    slope of $0.65\pm0.03$. The slope of the dotted line is 2/3.}
\end{figure}

We decompose the pattern propagation speed as 
$V_{\rm pattern}=V_{\rm radius}+V_{\rm shift}$
to distinguish the speeds corresponding to the center shift and the 
radius increase of the characteristic circles. The last four columns 
of Table\,\ref{tab:speed} list $V_{\rm radius}$ and $V_{\rm shift}$ 
for the outer and inner circles as the boundaries of the excess 
density regions. 
Figure\,\ref{fig:speed} illustrates our results in Table\,\ref{tab:speed}.
The speeds corresponding to the increase of radius for the outer and inner 
circles, $V_{\rm out,\,radius}$ and $V_{\rm int,\,radius}$, respectively, 
show a good correlation with the average of the intrinsic wind speed 
(slightly better than the average of the resulting wind speed). The 
speeds for the outer circles are approximately the average wind speed, 
while the speeds for the inner circles are smaller by as much as $2\cs$. 
A fit to the 10 points results in a slope of $1.04\pm0.03$. 
On the other hand, the centers of the circles shift with 
a speed proportional to the orbital velocity of the mass losing star. 
The speeds of the center shifts for the outer and inner circles are 
found to have a systematic difference of $0.5\cs$. Using the 10 points 
after correction of the systematic difference, a fit for the slope results 
in a value of $0.65\pm0.03$. Therefore, we take the propagation speeds in 
the radial direction for the outer and inner boundaries of the pattern as
\begin{equation}\label{eqn:voutc}
  V_{\rm out} = V_{w,0} + \frac{2}{3}V_p
\end{equation}
and
\begin{equation}\label{eqn:vintc}
  V_{\rm int} = \left(V_{w,0}-2\cs\right)
  + \frac{2}{3}\left(V_p-\frac{1}{2}\cs\right),
\end{equation}
respectively. The dashed lines in Figures \ref{fig:subso}(a)--(b),
\ref{fig:super}(a)--(b), and \ref{fig:fastw}(a)--(b), based on 
Equation (\ref{eqn:voutc}) combined with Equation (\ref{eqn:shape}), 
trace the density peaks at the outer edge of the pattern very well.

\subsubsection{Standoff}\label{sec:stand}

The innermost region of the shock emanates from a position which we 
denote as the standoff distance. An estimate for this shock location 
is analogous to earlier work of \citet{rag90}, who considered a one%
-dimensional accelerating jet in which one gas parcel can catch up 
the slower parcel ejected earlier. Since the required time for their 
encounter is the speed over the acceleration, the shock is located at 
a distance of (speed)$^2$/(acceleration). Based on this concept, 
\citet{rag11} proposed a simple analytical model to explain the shock 
standoff distance from the mass losing star in circular motion under 
the action of the centripetal acceleration, $V_p^2/r_p$. In the rest 
frame of the mass losing star, however, the ejected wind is not only 
affected by the centrifugal acceleration in the radial direction, but 
also by the Coriolis acceleration ($V_pV_{w,0}/r_p$), perpendicular 
to the instantaneous direction of the wind parcel. Hence, obtaining 
the standoff distance is no longer a one-dimensional problem. 

Nonetheless, we empirically find that this approach is a good 
approximation, as summarized in Table\,\ref{tab:stand}. In the 
second column of the table, the distance between the star and 
the first arm of the spiral pattern is listed. Here, we measure 
the distance to the shock in the direction opposite from the 
system center. In the cases that the wind and orbital speeds are 
comparable (models M2, M3, and M4), the shock standoff distance 
is comparable to \citeauthor{rag11}'s prediction (third column in 
Table\,\ref{tab:stand}). However, for models M1, M5, and M6, in 
which the wind is much faster than the orbital speed, significant 
differences are present. A modified formula,
\begin{equation}\label{eqn:stand}
  r_{\rm stand}/r_p = \left(\frac{2}{3}\frac{V_{w,0}+V_p}{V_p}\right)^2,
\end{equation}
is found to represent the standoff distance for all simulated data sets.

\subsubsection{Flattening}\label{sec:vflat}

As described in Section~\ref{sec:speed}, the positions for the 
density peaks in the meridional planes can be well traced by 
circular arcs. Furthermore, by comparing the structures displayed 
in Figure \ref{fig:subso}(b) and \ref{fig:super}(b), it is evident 
that a higher orbital motion of the mass losing star causes the 
density distribution of the resulting circumstellar envelope to be 
more oblate-shaped. Here, we use the shape of the circular arcs to 
determine the overall vertical shape of the circumstellar envelope. 

Using Equation (\ref{eqn:voutc}), we define the eccentricity of the 
vertical shape as $e=(a^2-b^2)^{1/2}/a$, where $a=V_{w,0}+2V_p/3$ 
and $b=V_{w,0}$. The eccentricities for models M1--M6 are 0.42, 0.72, 
0.81, 0.63, 0.43, and 0.47, showing that the model characterized by a 
relatively large orbital-to-wind velocity ratio (i.e., models M2--M4) 
has a more flattened shape. Hence, the overall eccentric shape of an 
observed pattern can be used to constrain the orbital velocity of the 
mass losing star with a given wind velocity. 

\begin{figure}
  \plotone{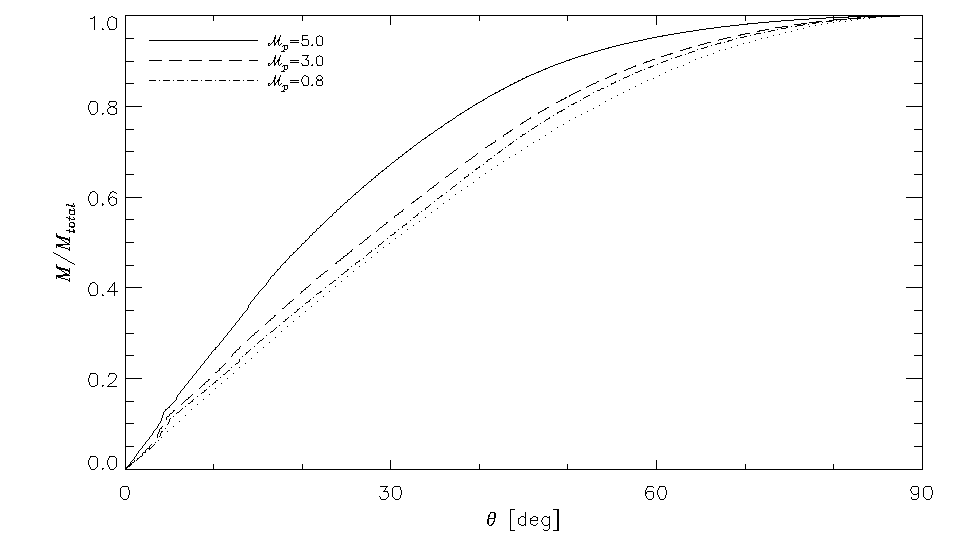}
  \caption{\label{fig:vflat}
    Integrated mass normalized by the total gas mass as a function 
    of the latitudinal angle from the orbital plane. The models M1 
    ($\mach_p=0.8$; dot-dashed), M2 ($\mach_p=3.0$; dashed), and M3 
    ($\mach_p=5.0$; solid) are compared to show the dependence on 
    the orbital speed for a fixed wind model. For reference, the
    dotted line exhibits the sinusoidal function for the perfectly
    uniform density distribution.
  }
\end{figure}

Figure\,\ref{fig:vflat} exhibits another measure to quantify the 
flattening of the circumstellar envelope obtained by integrating 
the mass as a function of the angle from the orbital plane. 
For the fast orbital motion cases (M1--M3), the integrated mass 
normalized by the total mass shows a considerable systematic 
change with the orbital velocity of the mass losing star. This 
result indicates that the mass of the circumstellar envelope is 
concentrated toward the orbital plane, not just locally modulated 
within the pattern. 
This flattening is due to the action of the centrifugal force 
imparted to the gas by the motion of the mass losing star. As
the models with faster winds (models M4--M6) follow the curve 
for a uniform density distribution, the flattening effect as a 
function of the orbital velocity is not distinguishable. We have 
also checked that its dependencies on other parameters such as the 
mass and orbital radius of the mass losing star are not significant.

Previously, \citet{hug09} used the binary simulations of \citet{mas99} to 
find that the flatness in the density distribution is determined by a single 
parameter, $GM_{\rm comp}(r_p+r_{\rm comp})^{-1}(V_w(r_{\rm comp}))^{-2}$, 
with little dependence on $1+q$ where $q=M_p/M_{\rm comp}$. These results 
led them to conclude that the flattening of a circumstellar envelope in 
a binary system is caused by the gravitational focusing of the wind onto
the companion. However, their first parameter can be rewritten as 
$(V_p/V_w)^2\times(1+q)$ using the binary mass ratio $q$ factor. 
Given that the flattening is insensitive to $q$, we find that their 
flattening parameter can be simplified to $(V_p/V_w)^2$, which is 
consistent with the tendency found from our simulation results.

\subsubsection{Inclination}\label{sec:incli}

\begin{figure}
  \plotone{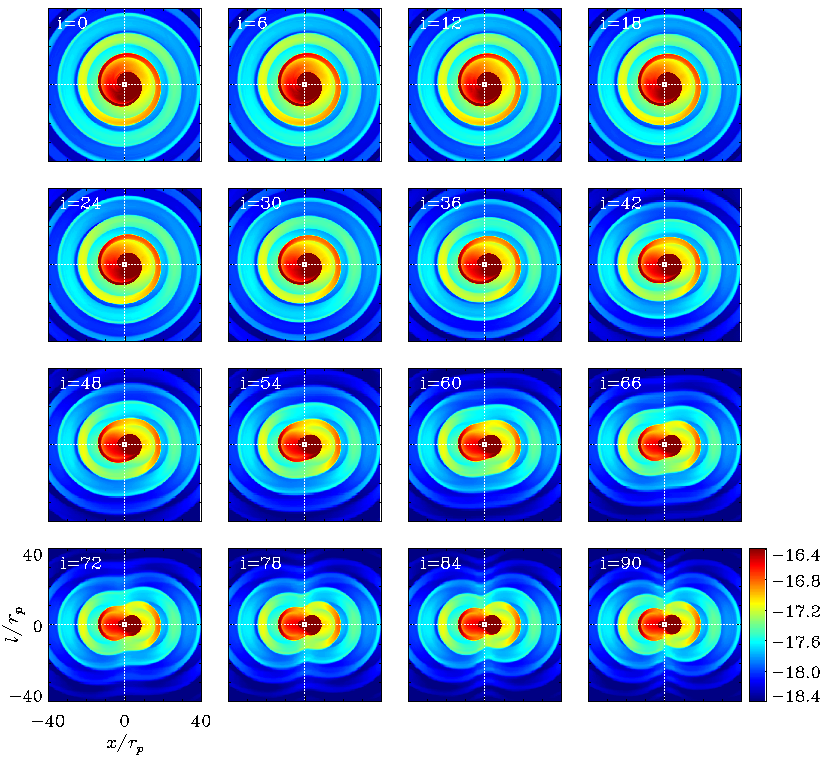}
  \caption{\label{fig:incli}
    Density distribution maps for model M3 as a function of the
    viewing inclination angle $i$. The vertical axis is defined
    by $l=y\cos i+z\sin i$, where $z$ represents the orbital axis.
    All lengths are scaled to the orbital radius $r_p$ of the
    mass losing star about the center of mass. Color bar labels
    the density in unit of $g$\,cm$^{-3}$ in logarithmic scale.
  }
\end{figure}

The three-dimensional structure of the circumstellar envelope 
for a given model exhibits various shapes in the plane of the 
sky depending on the viewing inclination, from a spiral (0\arcdeg; 
face-on) to ring-like pattern (90\arcdeg; edge-on). For example, 
the case shown in Figure\,\ref{fig:incli}, corresponding to a model 
in which the orbital and wind velocities are comparable, reveals a 
peanut-shaped pattern when viewed edge-on because of the horizontal 
shifts of the ring-like pattern as described in Section \ref{sec:speed}.

The appearance of the patterns is spiral-like for a wide range 
of viewing angles, excluding angles very close to the edge-on 
($\gtrsim$\,70\arcdeg--80\arcdeg). Previously, \citet{he07} 
pointed out that the majority of observed circumstellar patterns 
around evolved stellar binary systems will appear spiral-like. 
However, he only explored the case of a very fast wind compared 
to the orbital speed (the ratio greater than 10) in a simple piston 
model. Hence, the systematic elongation in the overall shapes of 
the spirals was not found, which is prominent in, for example, 
our model M3 shown in Figure\,\ref{fig:incli}. 

We also find that the elongated spirals in the mid-range of viewing 
inclinations are tilted. That is, the line connecting the longest axis 
of the elongated spiral in the plot for $i=60$\arcdeg, for example, 
does not match with the $l=0$ axis, which is the intersection of the 
viewing plane with the orbital plane. Here, the orientation of the 
tilt depends on the orientation of the orbital motion. 

\begin{figure}
  \plotone{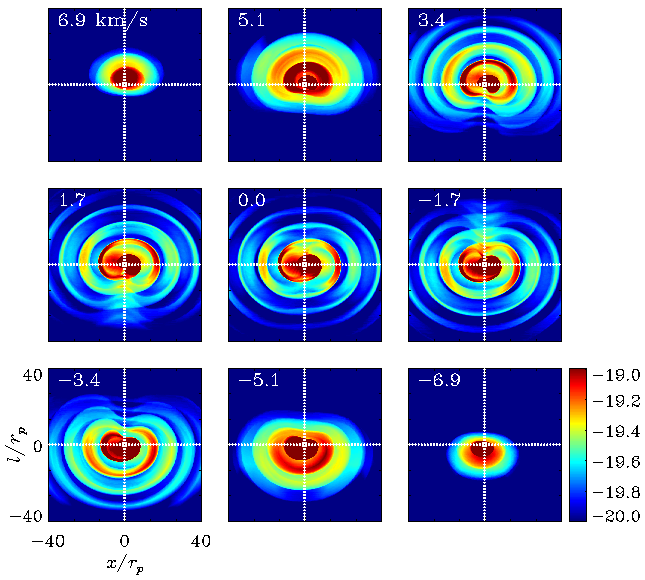}
  \caption{\label{fig:chmap}
    Velocity channel map for model M3 with a viewing inclination 
    of 60\arcdeg. The color bar refers to the integrated density 
    normalized by the total range of velocity (15.4\kmps) and the 
    domain size (800\,AU) in unit of $g$\,cm$^{-3}$ in logarithmic 
    scale. The central velocity of each channel is denoted on the 
    top left of each panel. The channel width is 1.7\,\kmps.
  }
\end{figure}

Figure\,\ref{fig:incli} shows the density distribution in the mid plane 
in the line-of-sight direction, which dominates the central channel of 
a velocity channel map (see middle panel of Figure\,\ref{fig:chmap}). 
In different velocity channels, however, the patterns exhibit different
tilt orientations and angles. The middle row of Figure\,\ref{fig:chmap} 
demonstrates the channels close to the central velocity, reflecting 
the density distribution close to the mid plane. The longest axes of 
the patterns in these three panels are rotated counterclockwise from 
$l=0$ axis. However, in the outer layers (or in higher velocity channels), 
the patterns are twisted and the longest axes are rotated in the clockwise 
direction. This trend reverses when the star orbits in the clockwise 
direction, opposite to the cases displayed here. It is also worth noting 
that the column density map does not show the mismatch of the axis of 
the pattern from the line of nodes since all the channels characterized 
by different degrees of mismatch are integrated out. 

\subsection{Effects of Companion Wake}\label{sec:2arms}

In previous sections, we focused on the asymmetric pattern caused by 
the reflex motion of the mass losing star as a result of the indirect 
influence of the binary companion. In those models, we suppressed the 
direct effect of the companion, which assembles wind material to create 
a gravitational wake. In model M7, we include the influences of both 
stellar motions in modulating the circumstellar envelope around an 
evolved binary stellar system. By comparing this model with model M6, 
whose initial setup is the same as M7 except for the gravitational 
wake of the companion star, we separately investigate the direct and 
indirect effects due to a companion star. 

The properties of a companion's gravitational wake is described by 
\citetalias{kim12} in detail. In this previous work, the companion 
is assumed to be a substellar mass object in order to minimize the 
reflex motion of the mass losing star. Although the companion's mass 
adopted in this paper is comparable to the AGB star, we show that 
the previous results can be applied in this case as well.

Similar to the pattern caused by the motion of the mass losing star, 
the gravitational wake of a companion described in \citetalias{kim12} 
exhibits a single-armed spiral in the orbital plane and a circular arc 
pattern in meridional planes. However, these two patterns have three 
differences, which complicate the resulting structures. 

The first difference is that the companion's wake is attached to the 
companion\footnote{For a massive companion, the gravitational wake 
can be detached by up to distance of $GM_{\rm comp}(V_{\rm comp}^2-
\cs^2)^{-1}r_s^{-1}$ in a static background \citep[see][]{kim09,kwt10}. 
However, this detachment hardly occurs in outflowing background because 
the material is swept out by the wind \citepalias{kim12}. Here, $r_s$ 
is the gravitational softening radius of the object.}, while the pattern 
due to the motion of the mass losing star stands off at the distance as 
in Equation (\ref{eqn:stand}). 
The second difference stems from the propagation speeds of the patterns 
determining the shape. For the companion's wake, the centers of the arcs 
appearing in meridional planes are fixed at the orbital distance of the 
companion, i.e., $V_{\rm shift}=0$ \citep[see for example Figure\,1 of]
[]{kim11} and the radii of the arcs increase with speed $V_{\rm radius}
=V_w\pm\cs$, where the upper and lower signs correspond to the outer and 
inner edges of the regions of excess density, respectively (c.f., Equations 
(\ref{eqn:voutc})--(\ref{eqn:vintc})). 
Third, while the arc pattern due to the reflex motion of the mass losing 
star nearly reaches the orbital axis, the gravitational wake of the companion 
forms within a very limited height from the orbital plane, up to the angle of 
$\tan^{-1}((\mach_{\rm comp}^2-1)^{1/2}/(1+0.2\mach_w\mach_{\rm comp}))$,
that is Equation (13) from \citetalias{kim12}. 

\begin{figure}
  \epsscale{0.8}
  \plotone{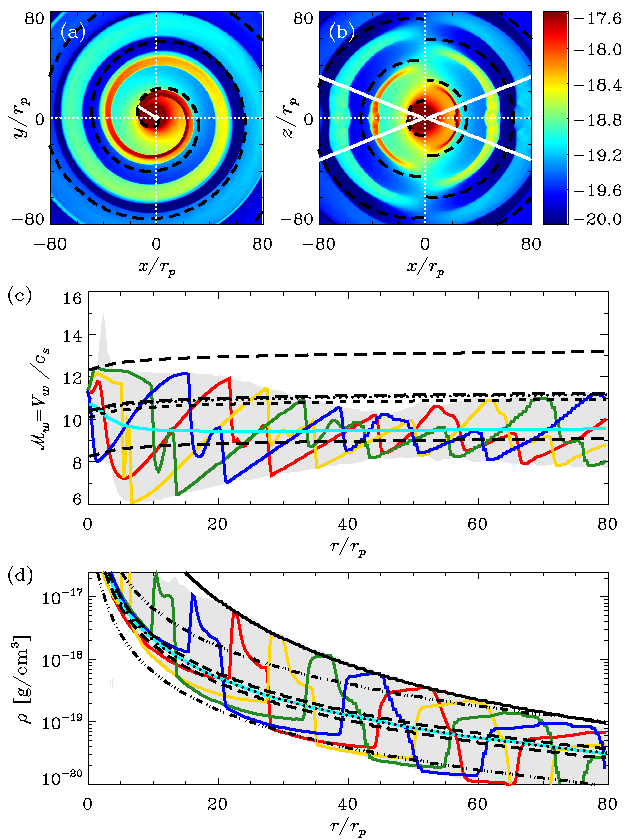}
  \caption{\label{fig:2arms}
    Same as Figure~\ref{fig:fastw}, but for model M7 including the 
    direct effect of the companion forming a gravitational wake.
    The dashed lines in (a) and (b) refer to the lines in Figure 
    \ref{fig:fastw}(a)--(b), for comparison with the shapes 
    in model M6 excluding the companion's direct effect. Note that
    the shape of the outer shock can be well formulated by, in the 
    pattern propagation speed (Eq.\ (\ref{eqn:voutc})), replacing 
    $V_{w,0}$ (dotted line in (c)) by $V_w$ as the average profile 
    of actual wind velocity. The white line in (b) is the vertical
    extension limit of the companion wake \citepalias[See][]{kim12}. 
    The black solid line in (d) shows the linear superposition of 
    the maximum density fluctuations caused by the orbital motions 
    of the individual stars.
  }
\end{figure}

In Figure \ref{fig:2arms}(a)--(b), we present the shape of pattern 
in model M7 for the binary system, overlaid with the corresponding 
shape in model M6 excluding the gravitational wake of the companion 
(dashed lines). It is definitely seen that the pattern shape has 
contracted by including the direct effect of the companion. This 
difference arises due to the existence of a shock front producing 
a distinctive inner edge of the spiral arm structure in Figure 
\ref{fig:2arms}(a), which was absent in Figure \ref{fig:fastw}(a) 
without the companion's wake.

The flow slows immediately after the spiral shock corresponding to the 
inner edge of the companion wake. For instance, the red line in Figure 
\ref{fig:2arms}(c) for the fluid speed along $^+x$-axis drops at 
$r/r_p\sim22$ and $\sim27$. In contrast, in Figure \ref{fig:fastw}(c) 
without the companion wake, it decreases continuously in the range of 
$r/r_p\sim24$--29. Since the flow after the shock at $r/r_p\sim22$ in 
model M7 is slower than in the case where the companion wake is not 
taken into account, the pattern propagation speed is decreased thus 
shrinking the pattern.

This shape discrepancy between the models with (M7) and without (M6) 
the gravitational wake of the companion is, however, insignificant at 
high latitudes as shown in Figure \ref{fig:2arms}(b). This latitude 
dependence is related to the white solid lines in the 
figure, which indicates the vertical extension limit of the companion's 
wake influence, from \citetalias{kim12}. Indeed, one can easily find the 
clumpy structures within the limits, which is the wake of the companion. 
Since the shock associated with the companion's wake does not exist 
at high latitudes, the wind can flow in a similar manner to the case 
excluding the companion's wake effect.

\begin{figure}
  \epsscale{1}
  \plotone{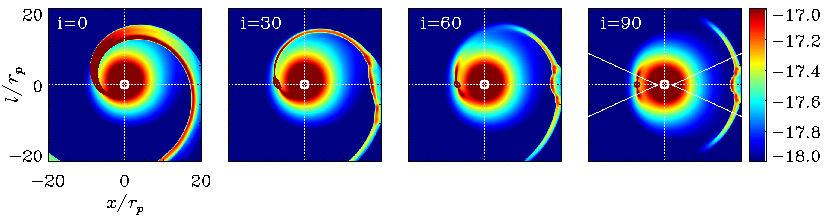}
  \caption{\label{fig:clump}
    Density distribution in the midplane of the sky for model M7 
    with the viewing inclination angles of 0\arcdeg, 30\arcdeg, 
    60\arcdeg, and 90\arcdeg. The gravitational density wake of 
    companion overlapped with the structure due to the reflex 
    motion of the mass losing star creates the sub-arm structures 
    creating a dense inner arm in the face-on view, but appearing 
    as clumpy structures in the inclined views. The white lines on 
    the rightmost panel display the vertical extension limit of the 
    companion wake from \citepalias{kim12}. The white thick circles 
    represent the orbital radius $r_p$ of the mass losing star, and 
    the companion is orbiting at $2r_p$. The vertical axis shows the
    projected distance, $l=y\cos i+z\sin i$, where $z$ represents the 
    orbital axis. The density in unit of $g$\,cm$^{-3}$ is color-coded
    in logarithmic scale.
  }
\end{figure}

\begin{figure}
  \plotone{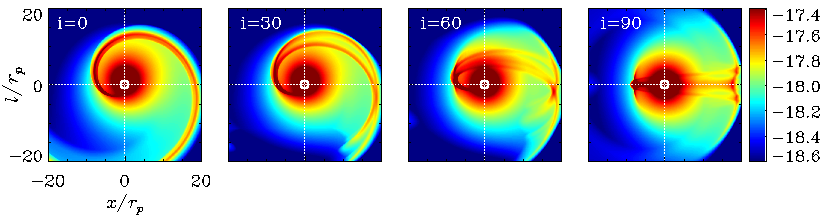}
  \caption{\label{fig:colmn}
    Same as Figure~\ref{fig:clump}, but for the average density
    calculated by the column density over the domain size.
  }
\end{figure}

The contribution of a companion's wake to the structure due to the 
motion of the mass losing star is clearly seen in a zoomed image in 
Figure\,\ref{fig:clump}. The plots for inclinations $i>0$\arcdeg\ 
display clumpy structures due to the overlap of the companion's 
wake, while for the face-on inclination a continuous, distinct 
inner edge of the spiral arm is seen. 
In the panel for $i=30$\arcdeg, the wake of the companion presents 
a distinctive knot at $x/r_p\sim-8$ and a wiggly distribution 
on the opposite side connected by a spiral pattern due to the 
primary star's reflex motion. It is worth noting that the spiral 
is patchy for an inclination $i=60$\arcdeg. We emphasize that the 
knots do not represent the companion star and the spiral patterns 
in these inclinations are not the companion's wake. The latter 
point is clarified by the corresponding column density in Figure%
\,\ref{fig:colmn}, revealing that the knotty structures are indeed 
due to a superposition of two spiral patterns formed by the motions 
of the individual stars.

In contrast to the structure due to the reflex motion of the mass 
losing star, the wake of the companion star significantly changes 
its apparent shape as a function of the inclination angle
(Figure\,\ref{fig:colmn}). This is due to the fact that the 
companion wake is vertically thin, so that it closely corresponds 
to the orbital plane. Therefore, if the emission is sufficiently 
optically thin, but sensitive enough, to detect the companion wake, 
the viewing inclination angle may be inferred from the positional 
displacement between the two spiral-like structures.

From the black solid curve in Figure \ref{fig:2arms}(d), we clarify 
that the density enhancement in the overlap regions with respect to 
the average density (calculated by $(\rho_{\rm max}-\rho_0)/\rho_0$) 
is consistent with a linear superposition of the amounts of density 
enhancements due to the two mechanisms. To account for the effect of 
reflex motion to the density enhancement in Figure \ref{fig:fastw}(d), 
we have used a fitting formula (given in Table\,\ref{tab:dfits}) for 
the maximum density enhancement as a function of distance. For the 
calculation of the density enhancement in the gravitational wake 
of the companion, we have taken the formula\footnote{Equation (12) 
of \citetalias{kim12} is rewritten as \begin{displaymath}
  \delta\rho/\rho=\frac{GM_{\rm comp}/\cs^2}{|r-r_{\rm comp}|}
  \left(\frac{\frac{|r-r_{\rm comp}|}{GM_{\rm comp}/\cs^2}V_w^2+
    10V_wV_{\rm comp}+\cs^2}{V_w^2+|V_{\rm comp}^2-\cs^2|}\right)^{1/2}.
\end{displaymath}} empirically found in \citetalias{kim12}. The 
summation of these 
enhancement values (black solid line in the figure) matches very well 
with the maxima of the density profiles (red, yellow, green, and blue 
colors) in the orbital plane of the binary system. The change in minimum 
densities from the values in Figure \ref{fig:fastw}(d) is negligible 
\citepalias[see Figures 5--6 in][]{kim12}.

\section{DISCUSSION}\label{sec:discu}

As shown in the previous sections, the motion of the components in 
a binary system can generate various morphologies in the outflowing 
circumstellar envelopes of stars in the AGB phase. A detailed examination 
of the shape of the pattern, the density profile, and the kinematics can, 
in principle, provide constraints on the binary orbital properties. In the 
following, we apply our theoretical results to provide an interpretative 
framework of the large scale global properties of the observed asymmetric 
pattern in the circumstellar envelope of the carbon-rich AGB star AFGL 3068.

AFGL 3068 is one of the best test cases for investigation within the 
framework of the binary scenario since it possesses a well-defined 
spiral pattern extending over a large area of the circumstellar envelope. 
It is seen in optical wavelengths \citep[][\citetalias{mau06} hereafter]
{mau06} as a result of the scattering by dust illuminated by Galactic 
light. Observational evidence exists for two point-like, near-infrared 
sources at the center \citep{mor06} of the envelope with a projected 
separation of $0.\!\!\arcsec11$ (109 AU, assuming the distance of 1 kpc). 
Five turns of the spiral pattern are detected over 12\arcsec\ from the 
center ($\sim10^4$\,AU). \citetalias{mau06} used the spacing between 
spiral arms of 2.$\!\!$\arcsec29 (2300 AU) to derive a binary orbital 
period of 830 yr, assuming that the observed spiral pattern propagates 
outward with the speed (14\,\kmps) taken as the expansion velocity of 
the envelope.

From Figure\,8 of \citetalias{mau06}, however, we find that their sketch 
of an Archimedes spiral mismatches with the position of peak intensity. 
The observed pattern is elongated in the northwest-southeast direction. 
The ratio between the longest axis of the observed pattern and the length 
perpendicular to this axis is found to be 1.1. Considering that the
appearance of the pattern is unlikely in an edge-on view, the ratio in 
the edge-on view (or the aspect ratio of the oblate spheroid) must 
be larger ($a/b\geq1.1$). The corresponding eccentricity ($e\geq0.42$) 
is consistent with M4--M6 models, which correspond to the cases for an 
orbital velocity greater than 2.2\,\kmps\ with a similar wind velocity. 
Using the analysis in Section \ref{sec:vflat}, we obtain the velocity 
ratio of $V_p/V_{w,0}\geq0.15$.

An additional constraint on the orbital velocity of the mass losing star 
is found from the maximum line-of-sight velocity of 14\,\kmps. As seen 
in Figures \ref{fig:subso}(c), \ref{fig:super}(c), \ref{fig:fastw}(c), 
and \ref{fig:2arms}(c), the wind velocity cannot exceed $V_{w,0}+V_p$ in 
any direction, thus $V_{w,0}+V_p\ga14$\,\kmps. This velocity is not likely 
to be very different from 14\,\kmps. Combining this with the constraint 
derived from the eccentricity of the overall shape of the pattern yields 
$V_p\ga1.8$\,\kmps\ and $V_{w,0}$ close to 12.2\,\kmps. Models M4 
and M5 satisfy the conditions for the wind and orbital velocities.

The pattern spacing of $2.\!\!\arcsec29$ with the corrected pattern 
propagation speed of $V_{w,0}+2V_p/3$ yields an orbital period $T_p$ 
of about 820\,yr, which is not significantly different from 
\citetalias{mau06}'s result. The corresponding orbital radius 
of the mass losing star is $r_p\geq0.\!\!\arcsec050$ or 50 AU. 
Comparing this value with the {\it projected} binary separation 
of 109\,AU \citep{mor06} implies that either the binary mass ratio 
$q=M_p/M_{\rm comp}$ corresponding to $r_{\rm comp}/r_p$ is small 
(possibly less than 1) or the projection effect is significant (i.e., 
high inclination). With the binary separation of $\geq109$\,AU and 
the derived limit of the orbital velocity $V_p\ga1.8$\,\kmps, the 
binary mass ratio is constrained to be\footnote{It is calculated 
by $V_p^2r_p^{-1}=GM_{\rm comp}(r+r_{\rm comp})^{-2}$, or equivalently 
$q(1+q)=GM_p(r_p+r_{\rm comp})^{-1}V_p^{-2}\la8.5$--11.3.} $q\la2.5$%
--2.9, adopting the mass of a carbon star as $M_p\leq3$--4\,\Msun\ 
assuming such stars do not undergo the hot bottom burning process 
\citep[and references therein]{gar06}. 

The arm width of $\la0.\!\!\arcsec5$ reported in \citetalias{mau06} 
can also be used to infer a property of the circumstellar envelope 
when compared with the theoretical distance of double peaks (inner and 
outer edges of the arm). Equations (\ref{eqn:voutc})--(\ref{eqn:vintc}) 
indicate that the arm width corresponds to $2.34\cs\times T_p$, implying 
a sound speed $\cs\la1.2$\,\kmps\ or temperature less than 150 K, if 
$T_p=820$\,yr. We note that our M4 model assumes a sound speed of 1\,\kmps, 
which is in the range estimated from the observed arm width. The velocity 
spread corresponding to broadening of the arm width ($2.34\cs$) is 
small compared to the pattern propagation speeds for the outer and 
inner peaks, which prevents the overlap of the outer and inner peaks 
in the region where the scattered light is detected. However, molecular 
line observations indicate that the temperature of the circumstellar 
envelope is plausibly much lower than 150 K at the distance where
the spiral pattern is detected \citep[10--30\,K,][]{woo03}. The 
difference in temperature may imply that the arm width is not fully 
resolved within $0.\!\!\arcsec5$ or that our assumption of isothermality 
breaks down. Given that the isothermal assumption provides a very good 
description of the overall shape of the pattern, the broadening of the 
arm width may be caused by a local temperature increase within the 
shocked spiral arm. 

In a separate paper, \citet{submt} \oops{modeled the circumstellar envelope 
of AFGL 3068 using an adiabatic equation of state with the specific 
heat ratio of 1.4. The calculations show that the temperatures in the 
shocked arm are as high as 1000\,K. In the range of arm temperatures 
($\sim$\,100--1000\,K), the heating timescale due to the gas-dust 
collisions in the postshock region for 0.1\,$\micron$-sized grains is 
similar to (or slightly smaller than) the timescale for cooling via 
molecular line transitions \citep{edg08}, indicating that the 
adiabatic equation of state without including the above heating and 
cooling mechanisms should be reasonable. Here, we note the important 
point that the spiral shape for a non-isothermal gas is as predicted 
in Equation (\ref{eqn:voutc}) of the current work}.

Finally, we compare the theoretical prediction for the density contrast 
with the value estimated from the observations. From the intensity jump 
of a factor of up to $\sim2.5$, \citetalias{mau06} derived the density 
contrast corresponding to a factor of up to $\sim5$ assuming a nested 
shell model. In comparison, our spiral model predicts a density 
contrast up to $\sim(1+\mach_p)^2\ga6$ with $V_p\ga1.8$\,\kmps\ and 
$\cs\la1.2$\,\kmps. The contrast decreases with increasing distance 
from the center, as low as $(1+\beta_0)^2$ (a level of 1.3) beyond 
the distance where overlap of the inner and outer shocks occurs. 
The contrast in the observed intensity profile may correspond to 
the contrast of column density if the scattering is optically thin. 
We note that the first arm in model M4 is characterized by a column 
density contrast of a factor of 2--7 and the volume density contrast 
as high as a factor of a few tens, depending on the distance from the 
center as well as the choice of the reference value.

\section{SUMMARY AND CONCLUSION}\label{sec:concl}

We have carried out three dimensional hydrodynamic simulations to 
investigate the effects of a companion to an AGB star in a binary 
system to understand the physics and kinematics of the observed 
asymmetric patterns in outflowing circumstellar envelopes. It is 
found that the reflex motion of the mass losing star, due to the 
indirect effect of the companion, dominates the overall structure 
of the circumstellar envelope. The direct effect associated with 
the gravitational wake of the companion results in a similar, but 
narrower pattern, imprinted on the broader spiral pattern. The 
overlaps of the two patterns result in the generation of clumpy 
structures. The appearance of these structures, as well as the 
spiral-like pattern, is evident for a wide range of inclination 
angles. By extending the study to include cases for a slow wind, 
we find a systematic flattening of the spiral-like pattern with the 
viewing inclination. A quantitative description of the flattening is 
presented in terms of a ratio between the wind and orbital velocities. 
In particular, constraints on the orbital speed of the mass losing star 
as well as the inclination angle of the system (from the flattening 
of the overall shape and the deviation of the detailed pattern from 
a spiral) can, in principle, be obtained. However, a high sensitivity%
/resolution observation is required to trace the full spiral pattern, 
given that the inclined spirals are possibly patchy. 
Additional constraints on the system can be obtained from the amplitudes 
of the density and velocity modulations as compared to those characteristic 
for a single mass losing star. The numerical results have been applied to 
the extreme carbon star AFGL 3068, and we find that the physical parameters 
are within reasonable ranges for circumstellar envelopes. 

Although our model comparison is consistent with the observed pattern 
shape and density contrast of AFGL 3068, several aspects can be improved 
in future work. For example, the observed arm width is thin compared to 
the modeled column density distribution, perhaps, implying that the sound 
speed is less than 1\,\kmps. In addition, the applied model does not match 
the small scale details, for which fine tuning of the model is required. 
Inclusion of the gravitational wake of the companion will be necessary as 
the companion mass sensitively affects the small scale details as well as 
the mass ratio in determining most of the orbital parameters in a binary 
system. Finally, the possibility of eccentric orbits should be considered 
as the departure from circular motion may modify the fine structures as 
suggested by \citet{he07} in a simple ballistic model.

\acknowledgments
We are grateful to the anonymous referee and Dr.\ N.~Mauron for 
their fruitful comments which helped to increase the significance 
and potential of this work.
This research is supported by the Theoretical Institute for Advanced 
Research in Astrophysics (TIARA) in the Academia Sinica Institute of 
Astronomy and Astrophysics (ASIAA). The computations presented here 
have been performed through the ASIAA/TIARA computing resource, using 
FLASH3.0 code developed by the DOE-supported ASC/Alliance Center for 
Astrophysical Thermonuclear Flashes at the University of Chicago.

\clearpage
\begin{deluxetable}{clcccclrcc}
\tablecolumns{10}
\tablecaption{Parameters for hydrodynamic simulations\label{tab:param}}

\tablehead{\colhead{} & \colhead{}
 & \multicolumn{4}{c}{Properties of Binary System} & \colhead{}
 & \multicolumn{3}{c}{Domain Information} \\
\cline{3-6} \cline{8-10} \\[-2ex]
\colhead{Model} & \colhead{}
 & \colhead{$V_p$} & \colhead{$V_w^\prime\ \ @\ \ r_*$} & \colhead{$\cs$}
 & \colhead{$M_p/M_{\rm comp}$\tablenotemark{\dag}}  & \colhead{}
 & \colhead{$L$} & \colhead{$\Delta L$}
 & \colhead{refinement\tablenotemark{\ddag}} \\
\colhead{} & \colhead{}
 & \colhead{[\kmps]} & \colhead{[\kmps] ([AU])} & \colhead{[\kmps]}
 & \colhead{} & \colhead{} & \colhead{[AU]} & \colhead{[AU]} & \colhead{}}
\startdata
M1 %
 && 0.8 & 0.1 (2) & 1 & (4.6) && 1600 & 0.2 -- 12.5 & 9 -- 3 \\
M2 %
 && 3.0 & 0.1 (2) & 1 & (1.5) &&  400 & 0.2 -- 3.13 & 7 -- 3 \\
M3 %
 && 5.0 & 0.1 (2) & 1 & (1.0) &&  400 & 0.2 -- 3.13 & 7 -- 3 \\
M4 %
 && 4.7 & 10. (2) & 1 & (1.0) && 1600 & 0.2 -- 12.5 & 9 -- 3 \\
M5 %
 && 2.2 & 10. (2) & 2 & (2.0) && 1600 & 0.2 -- 12.5 & 9 -- 3 \\
M6 %
 && 2.2 & 10. (2) & 1 & (2.0) && 1600 & 0.2 -- 12.5 & 9 -- 3 \\
M7 %
 && 2.2 & 10. (2) & 1 &  2.0  && 1600 & 0.2 -- 12.5 & 9 -- 3 \\
\enddata
\tablecomments{Each column represents 
[1] model name, 
[2] orbital speed of the mass losing star, 
[3] wind launching speed on the stellar surface, defined by the 
boundary of the resetting radius ($r_*$),
[4] isothermal sound speed, 
[5] mass ratio between the mass losing star and the companion, 
[6] half of domain size, 
[7] range of spatial resolution, 
and
[8] corresponding range of refinement level with the mother grids of 
64$\times$64$\times$32 in $(x,y,z)$-directions. Mass $M_p$, mass loss 
rate $\dot{M}_p$, and orbital radius $r_p$ of the mass losing star are 
1\,\Msun, $10^{-6}$\,\Mspy, and 10 AU, respectively.
}
\tablenotetext{\dag}{Direct effect of the gravitational potential of 
the companion is included only in model M7. In the other models, the
companion's gravitational wake is suppressed so that the only role of 
companion is to cause the mass losing star to orbit around the center 
of mass.}
\tablenotetext{\ddag}{Static, centrally concentric mesh is employed.}
\end{deluxetable}

\begin{deluxetable}{clcclcc}
\tablecolumns{7}
\tablecaption{Maximum density profile\label{tab:dfits}}

\tablehead{\colhead{Model} & \colhead{}
 & \multicolumn{2}{c}{$\rho_{\rm max}$} & \colhead{}
 & \multicolumn{2}{c}{$(\rho_{\rm max}-\rho_0)/\rho_0$}\\
\cline{3-4} \cline{6-7}\\
 \colhead{} & \colhead{}
 & \colhead{$a\times10^{14}$} & \colhead{$n$} & \colhead{}
 & \colhead{$a$} & \colhead{$n$}}
\startdata
M1 &&  1.2 & $-2.3$ &&  5.1 & $-0.5$\\
M2 &&  2.5 & $-2.5$ &&   13 & $-0.7$\\
M3 &&  4.8 & $-2.8$ &&   33 & $-1.0$\\
M4 &&  6.8 & $-3.3$ && 1100 & $-1.8$\\
M5 && 0.39 & $-2.6$ &&  120 & $-1.1$\\
M6 &&  5.4 & $-3.1$ && 2300 & $-1.7$\\
\enddata
\tablecomments{$f(r)=a(r/r_p)^n$ is taken for the fitting functional form.
For all models, the slope of the mean density profile is measured to be in 
the range of $-2.0$ to $-2.2$ depending on distance from the orbital center.}
\end{deluxetable}

\begin{deluxetable}{clccclcclcc}
\tablecolumns{11}
\tablecaption{Pattern propagation speed\label{tab:speed}}

\tablehead{\colhead{Model} & \colhead{}
 & \colhead{$V_p$} & \colhead{$<V_{w,0}>$} & \colhead{$<V_w>$} & \colhead{}
 & \multicolumn{2}{c}{$\Delta{\rm(radius)}\,/\,T_p$} & \colhead{}
 & \multicolumn{2}{c}{$\Delta{\rm(shift)}\,/\,T_p$}\\
\cline{7-8} \cline{10-11}\\
\colhead{} & \colhead{} & \colhead{} & \colhead{} & \colhead{} & \colhead{}
 & \colhead{outer} & \colhead{inner} & \colhead{}
 & \colhead{outer} & \colhead{inner}}
\startdata
M1  && 0.8 & 5.2 & 5.3 \\
M2  && 3.0 & 4.7 & 5.2  && 5.1 & 1.8  && 2.1 & 1.8\\
M3  && 5.0 & 4.7 & 6.3  && 5.4 & 3.3  && 3.0 & 2.7\\
M4  && 4.7 & 11. & 12.  && 11. & 8.6  && 3.2 & 3.0\\
M5  && 2.2 & 14. & 14.  && 15. & 12.  && 1.1 & 0.4\\
M6  && 2.2 & 11. & 11.  && 11. & 9.9  && 1.4 & 1.0\\
\enddata
\tablecomments{Each column represents 
[1] model name, 
[2] orbital velocity of the mass losing star,
[3] average of the intrinsic wind velocity in the orbital plane 
over the simulated domain,
[4] average of the resulting wind velocity in the orbital plane 
over the simulated domain,
[5] and [6] increases of the radii of circles in $(x,z)$ plane for the 
outer and inner boundaries of the high density regions, respectively,
and [7] and [8] increases of the shifts of the centers of the circles.
Here, $T_p$ represents the orbital period.
}
\end{deluxetable}

\begin{deluxetable}{cccc}
\tablecolumns{4}
\tablecaption{Standoff distance\label{tab:stand}}

\tablehead{\colhead{Model} & \colhead{$r_{\rm stand}/r_p$}
 & \colhead{$(<V_{w,0}>/V_p)^2$} & \colhead{$(0.66\,(<V_{w,0}>+V_p)/V_p)^2$}}
\startdata
M1\tablenotemark{\dag} & 21. & 42. & 25.\\
M2\phm{\dag} & 2.5 & 2.5 & 3.0\\
M3\phm{\dag} & 1.4 & 0.9 & 1.7\\
M4\phm{\dag} & 4.9 & 5.6 & 5.0\\
M5\tablenotemark{\dag} & 22. & 39. & 23.\\
M6\phm{\dag} & 18. & 25. & 16.\\
\enddata
\tablenotetext{\dag}{Because M1 and M5 models do not have definitely 
defined shock features, we choose the position showing relatively high 
density gradient.
}
\end{deluxetable}

\begin{thebibliography}{}
\bibitem[Balick \& Frank(2002)]{bal02} Balick, B., \& Frank, A.\ 2002, \araa, 40, 439 
\bibitem[Bondi(1952)]{bon52} Bondi, H.\ 1952, \mnras, 112, 195 
\bibitem[Bondi \& Hoyle(1944)]{bon44} Bondi, H., \& Hoyle, F.\ 1944, \mnras, 104, 273 
\bibitem[Castro-Carrizo et al.(2010)]{cas10} Castro-Carrizo, A., Quintana-Lacaci, G., Neri, R., et al.\ 2010, \aap, 523, A59 
\bibitem[Darwin(1879)]{dar79} Darwin, G.~H.\ 1879, Proceedings of the Royal Society of London, 29, 168 
\bibitem[Decin et al.(2011)]{dec11} Decin, L., Royer, P., Cox, N.~L.~J., et al.\ 2011, \aap, 534, A1 
\bibitem[De Marco(2009)]{dem09} De Marco, O.\ 2009, \pasp, 121, 316 
\bibitem[de Val-Borro et al.(2009)]{dev09} de Val-Borro, M., Karovska, M., \& Sasselov, D.\ 2009, \apj, 700, 1148 
\bibitem[Dinh-V.-Trung \& Lim(2009)]{din09} Dinh-V.-Trung, \& Lim, J.\ 2009, \apj, 701, 292 
\bibitem[Edgar et al.(2008)]{edg08} Edgar, R.~G., Nordhaus, J., Blackman, E.~G., \& Frank, A.\ 2008, \apjl, 675, L101
\bibitem[Fong et al.(2006)]{fon06} Fong, D., Meixner, M., Sutton, E.~C., Zalucha, A., \& Welch, W.~J.\ 2006, \apj, 652, 1626 
\bibitem[Garc{\'{\i}}a-Hern{\'a}ndez et al.(2006)]{gar06} Garc{\'{\i}}a-Hern{\'a}ndez, D.~A., Garc{\'{\i}}a-Lario, P., Plez, B., et al.\ 2006, Science, 314, 1751 
\bibitem[He(2007)]{he07} He, J.~H.\ 2007, \aap, 467, 1081
\bibitem[Hoyle \& Lyttleton(1939)]{hoy39} Hoyle, F., \& Lyttleton, R.~A.\ 1939, Proceedings of the Cambridge Philosophical Society, 34, 405 
\bibitem[Huggins(2007)]{hug07} Huggins, P.~J.\ 2007, \apj, 663, 342
\bibitem[Huggins et al.(2009)]{hug09} Huggins, P.~J., Mauron, N., \& Wirth, E.~A.\ 2009, \mnras, 396, 1805 
\bibitem[Iben \& Livio(1993)]{ibe93} Iben, I., Jr., \& Livio, M.\ 1993, \pasp, 105, 1373 
\bibitem[Johnson \& Jones(1991)]{joh91} Johnson, J.~J., \& Jones, T.~J.\ 1991, \aj, 101, 1735 
\bibitem[Kim(2011)]{kim11} Kim, H.\ 2011, \apj, 739, 102
\bibitem[Kim \& Kim(2009)]{kim09} Kim, H., \& Kim, W.-T.\ 2009, \apj, 703, 1278
\bibitem[Kim \& Taam(2012a)]{kim12} Kim, H., \& Taam, R.~E.\ 2012a, \apj, 744, 136 \citepalias{kim12}
\bibitem[Kim \& Taam(2012b)]{submt} Kim, H., \& Taam, R.~E.\ 2012b, \apjl, submitted 
\bibitem[Kim(2010)]{kwt10} Kim, W.-T.\ 2010, \apj, 725, 1069
\bibitem[Mastrodemos \& Morris(1999)]{mas99} Mastrodemos, N., \& Morris, M.\ 1999, \apj, 523, 357 
\bibitem[Mauron \& Huggins(2006)]{mau06} Mauron, N., \& Huggins, P.~J.\ 2006, \aap, 452, 257 \citepalias{mau06}
\bibitem[Mayer et al.(2011)]{may11} Mayer, A., Jorissen, A., Kerschbaum, F., et al.\ 2011, \aap, 531, L4 
\bibitem[Miszalski et al.(2009)]{mis09} Miszalski, B., Acker, A., Parker, Q.~A., \& Moffat, A.~F.~J.\ 2009, \aap, 505, 249 
\bibitem[Mohamed \& Podsiadlowski(2011)]{moh11} Mohamed, S., \& Podsiadlowski, P.\ 2011, Why Galaxies Care about AGB Stars II: Shining Examples and Common Inhabitants, 445, 355 
\bibitem[Morris(1987)]{mor87} Morris, M.\ 1987, \pasp, 99, 1115 
\bibitem[Morris et al.(2006)]{mor06} Morris, M., Sahai, R., Matthews, K., Cheng, J., Lu, J., Claussen, M., \& S{\'a}nchez-Contreras, C.\ 2006, Planetary Nebulae in our Galaxy and Beyond, 234, 469 
\bibitem[Neri et al.(1998)]{ner98} Neri, R., Kahane, C., Lucas, R., Bujarrabal, V., \& Loup, C.\ 1998, \aaps, 130, 1 
\bibitem[Nordhaus \& Blackman(2006)]{nor06} Nordhaus, J., \& Blackman, E.~G.\ 2006, \mnras, 370, 2004 
\bibitem[Nordhaus et al.(2010)]{nor10} Nordhaus, J., Spiegel, D.~S., Ibgui, L., Goodman, J., \& Burrows, A.\ 2010, \mnras, 408, 631 
\bibitem[Pascoli \& Lahoche(2010)]{pas10} Pascoli, G., \& Lahoche, L.\ 2010, \pasp, 122, 1334 
\bibitem[Raga et al.(1990)]{rag90} Raga, A.~C., Binette, L., Canto, J., \& Calvet, N.\ 1990, \apj, 364, 601 
\bibitem[Raga et al.(2011)]{rag11} Raga, A.~C., Cant{\'o}, J., Esquivel, A., Huggins, P.~J., \& Mauron, N.\ 2011, in Asymmetric Planetary Nebulae 5 Conference, ed. A. A. Zijlstra, F. Lykou, I. McDonald, and E. Lagadec (Jodrell Bank Centre for Astrophysics, Manchester), 185
\bibitem[Raghavan et al.(2010)]{rag10} Raghavan, D., McAlister, H.~A., Henry, T.~J., et al.\ 2010, \apjs, 190, 1 
\bibitem[Schmidt et al.(2002)]{sch02} Schmidt, G.~D., Hines, D.~C., \& Swift, S.\ 2002, \apj, 576, 429 
\bibitem[Soker(1994)]{sok94} Soker, N.\ 1994, \mnras, 270, 774 \citepalias{sok94}
\bibitem[Soker(1996)]{sok96} Soker, N.\ 1996, \apjl, 460, L53 
\bibitem[Taam \& Sandquist(2000)]{taa00} Taam, R.~E., \& Sandquist, E.~L.\ 2000, \araa, 38, 113 
\bibitem[Theuns \& Jorissen(1993)]{the93} Theuns, T., \& Jorissen, A.\ 1993, \mnras, 265, 946 
\bibitem[Trammell et al.(1994)]{tra94} Trammell, S.~R., Dinerstein, H.~L., \& Goodrich, R.~W.\ 1994, \aj, 108, 984 
\bibitem[van Winckel(2003)]{van03} van Winckel, H.\ 2003, \araa, 41, 391 
\bibitem[Winters et al.(2000)]{win00} Winters, J.~M., Le Bertre, T., Jeong, K.~S., Helling, C., \& Sedlmayr, E.\ 2000, \aap, 361, 641 
\bibitem[Woods et al.(2003)]{woo03} Woods, P.~M., Sch{\"o}ier, F.~L., Nyman, L.-{\AA}., \& Olofsson, H.\ 2003, \aap, 402, 617
\end{thebibliography}
\end{document}